\documentclass{ckm}                 

\usepackage{txfonts}            
\confname{Workshop on the CKM Unitarity Triangle, IPPP Durham, April
  2003}

\title{Experimental update on the exclusive determination of $|V_{cb}|$}

\author{A. Oyanguren\addressmark{}}

\address[]{IFIC, Edificio Institutos de Investigaci\'on, Apdo. Correos 22085, 
Valencia, E-46071, Spain}

\begin{document}

\begin{abstract}
In this talk a review on the exclusive determination of $|V_{cb}|$
is presented. Updated values of this quantity obtained from 
$\overline {\rm B} \to {\rm D} \ell^- \overline \nu_\ell$ 
and $\overline {\rm B} \to {\rm D}^* \ell^- \overline \nu_\ell$ decays are given.  
New measurements from B-factories are expected to come soon and 
the main challenges to improve the accuracy on $|V_{cb}|$ extracted 
from these decays are discussed.  
\end{abstract}

\maketitle

\section{Introduction}
One way to determine the CKM matrix element $|V_{cb}|$ is by exclusively 
measuring the differential decay width of the  
$\overline {\rm B} \to {\rm D} \ell^- \overline \nu_\ell$ or
$\overline {\rm B} \to {\rm D}^* \ell^- \overline \nu_\ell$ processes as a 
function 
of the kinematic variable 
$w$,
the product of the four-velocities of the initial and final mesons. 
The la\-tter decay channel has the experimental advantage of a larger bran\-ching ratio 
and less background. Even more profitable is the fact that this decay 
does not suffer from helicity su\-ppression near $w=1$, the point 
where the charmed meson is produced at rest.
At this point the theo\-retical description is 
well controlled by the Heavy Quark E\-ffective Theory (HQET)~\cite{hqet} and 
$|V_{cb}|$ can be determined with higher accuracy.
Howe\-ver, measurements of $|V_{cb}|$ through 
$\overline {\rm B} \to {\rm D} \ell^- \overline \nu_\ell$ decays,
although more difficult to perform,
are also important to check the consistency of the theory.

\section{$|V_{cb}|$ from $\overline {\rm B} \to {\rm D}^* \ell^- 
\overline \nu_\ell$}
The differential decay width of the $\overline {\rm B} \to {\rm D}^* \ell^- 
\overline \nu_\ell$ process can be expressed as~\cite{hqet}:
\begin{equation}
\frac{d\Gamma(\overline {\rm B} \to {\rm D}^* \ell^- \overline \nu_\ell)}{dw}=
\frac{G_F^2|V_{cb}|^2}{48\pi^3}{{\cal K}_{D^*}(w){\cal F}^2_{D^*}(w)} 
\label{eq:widthds}
\end{equation}
where ${\cal K}_{D^*}(w)$ is a phase space function and ${\cal F}_{D^*}(w)$ is 
the form factor describing the $\overline {\rm B} \to {\rm D}^*$ transition.
The shape of the form factor cannot be predicted by the theory but it can 
be constrained by using dispersion relations~\cite{caprini}. 
HQET gives the normalization of ${\cal F}_{D^*}(w)$ at zero recoil ($w=1$), 
where the ${\rm B}$ and ${\rm D}^*$ wave functions are 
completely overlapped, to be unity. 
Taking into account $1/m_Q$ and QCD corrections
to the heavy quark limit, the normalization yields~\cite{ckm02}:
\begin{equation}
{\cal F}_{D^*}(1)=0.91 \pm 0.04. 
\label{eq:f1}
\end{equation}
If ${\cal K}_{D^*}(w)$ was not zero at $w=1$, $|V_{cb}|$ could be 
directly extracted from the measured differential decay width at this point. 
As the phase space vanishes in this region, the differential decay width 
has to be extrapo\-lated, the quality of the extrapolation 
depending on the quality of the reconstruction efficiency near to the zero 
recoil point. 

\subsection{Signal reconstruction}
Measurements of $|V_{cb}|$ by reconstructing 
$\overline {\rm B} \to {\rm D^*} \ell^- \overline \nu_\ell$ decays 
have been performed by ALEPH~\cite{aleph}, 
DELPHI [5,6]  
and OPAL~\cite{opal} collaborations from $Z \to b\overline b$ decays, 
and by CLEO~\cite{cleo} and BELLE~\cite{belle} collaborations, from 
${\rm B\overline B}$ pairs coming from $\Upsilon(4{\rm S})$ decays.
When the B meson is coming from a Z decay, a large boost is given to the 
B and to its decay products. Se\-condary vertices can be better determined
than in case of a $\Upsilon(4{\rm S})$ decay where the two B mesons are
produced practically at rest. 
On the contrary, the energy of a B coming from a Z cannot be so well 
determined as if it was coming from a $\Upsilon(4{\rm S})$ decay, and the 
resolution on the $w$ variable deteriorates. To measure 
the $\overline {\rm B} \to {\rm D^*} \ell^- \overline \nu_\ell$ differential 
decay width, leptons and ${\rm D}^*$ candidates are selected.
The ${\rm D}^*$ meson is reconstructed by its decay into a D meson and a 
soft pion, the latter is produced almost at rest in the ${\rm D}^*$ rest frame. 
The D meson can be reconstructed using several decay channels.  
In case of experiments working at the Z, the good vertex se\-paration 
allows the ${\rm D}^*$ to be inclusively reconstructed by mainly de\-tec\-ting 
the soft pion and few particles from the ${\rm D}^0$, thus increasing the 
available statistics.  
When the B is produced at rest, the soft pion cannot be so well detected and
the efficiency for charged pions decreases as $w$ goes to 1. 
This does not happen for ${\rm D}^{*0}\to {\rm D}^+\pi^0$ decays where the 
soft $\pi^0$ is identified by its decay into two photons. The CLEO 
collaboration analyzes this channel in addition to the 
${\rm D}^{*+}\to {\rm D}^0\pi^+$ one.

\subsection{Background}
The most difficult source of background 
in $\overline {\rm B}~\to~{\rm D^*}~\ell^- \overline \nu_\ell$ decays, 
is due to ${\rm D}^*$'s coming from excited ${\rm D}^{**}$ states.
Decay properties of resonant and non-resonant ${\rm D}^{**}$ decays 
are not well established yet and these decays introduce an 
important uncertainty in the determination of $|V_{cb}|$.
At the
$\Upsilon(4{\rm S})$ energy, kinematic variables such as the 
cosine bet\-ween the B and the ${\rm D}^*-$lepton system, which are obtained 
ma\-king use of the missing energy measurement and the beam energy as a 
constraint, can be used to eliminate ${\rm D}^{**}$ decays.
The contamination 
from ${\rm D}^{**}$ is larger at the Z energy and it is more difficult 
to se\-parate. Experiments use topological variables based on ver\-tex 
separation and charge correlation to identify this source of background.

\subsection{$|V_{cb}|$ measurements}

\begin{figure}
\hbox to\hsize{\hss
\includegraphics[width=1.\hsize]{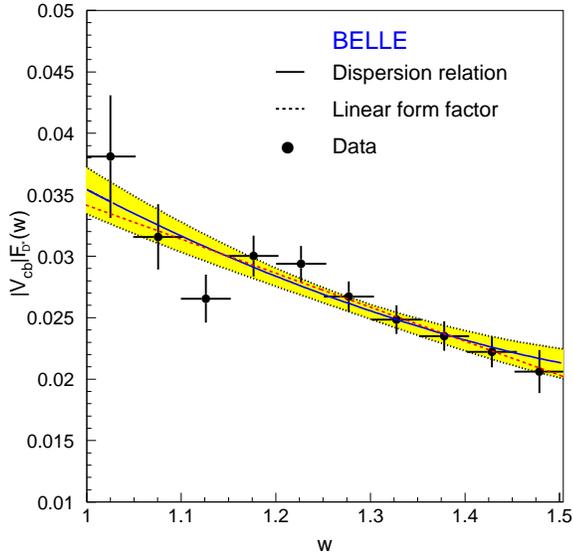}
\hss}
\caption{Unfolded distribution of ${\cal F}_{D^*}(w)|V_{cb}|$ as function 
 of $w$ measured by the BELLE collaboration.}
\label{fig:belle}
\end{figure}

The shape of the form factor ${\cal F}_{D^*}(w)$ entering in $d\Gamma/dw$ is 
usually expressed in terms of the form factor slope $\rho^2_{D^*}$ and 
of the form factor ratios $R_1$ and $R_2$~\cite{hqet}.
A parameterization of ${\cal F}_{D^*}(w)$, constrained by dispersion relations 
has been proposed in~\cite{caprini}.
The $R_1$ and $R_2$ form factor ratios have been measured by the CLEO 
collaboration~\cite{cleor1r2}. Experi\-ments use these values and fit $d\Gamma/dw$ to 
extract ${\cal F}_{D^*}(1)^2|V_{cb}|^2$ and the form factor slope $\rho^2_{D^*}$. 
Figure~\ref{fig:belle} shows the fit of the unfolded distribution of 
${\cal F}_{D^*}(w)|V_{cb}|$ measured by the BELLE 
collaboration~\cite{belle} using the
form factor of expression given in~\cite{caprini} or a linear parameterization. 

Averaging the results of the different experiments, 
the values of ${\cal F}_{D^*}(1)|V_{cb}|$ and $\rho^2_{D^*}$ have been 
found to be~\cite{hfag}: 

\begin{math}
\begin{array}{l}
{\cal F}_{D^*}(1)|V_{cb}|=(38.8\pm0.5(stat)\pm0.9(sys))\times10^{-3}\\
 \rho_{D^*}^2=1.49\pm0.05(stat)\pm0.14(sys). 
\end{array}
\end{math}

\begin{figure}
\hbox to\hsize{\hss
\includegraphics[width=1.\hsize]{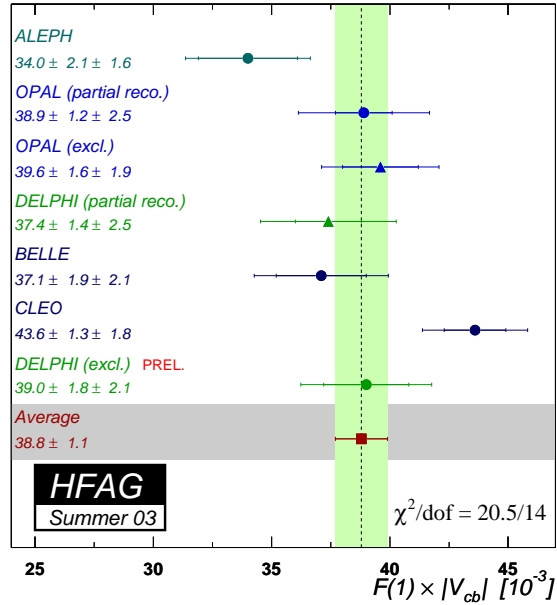}
\hss}
\caption{ The ${\cal F}_{D^*}(1)|V_{cb}|$ world average performed by the 
 Heavy Flavor Averaging Group \cite{hfag}, after rescaling the analyses to 
 common input parameters. The labels {\it 'partial reco.'} and {\it 'excl.'} 
 have 
 been used to
 distinguish analyses where the ${\rm D}^0$ is partially or exclusively 
 reconstructed, respectively. The label 'PREL.' refers to the preliminary result 
 of \cite{delphi_prel}. Only published results are taken into account in the average.}
\label{fig:average}
\end{figure}

\begin{figure}
\hbox to\hsize{\hss
\includegraphics[width=1.\hsize]{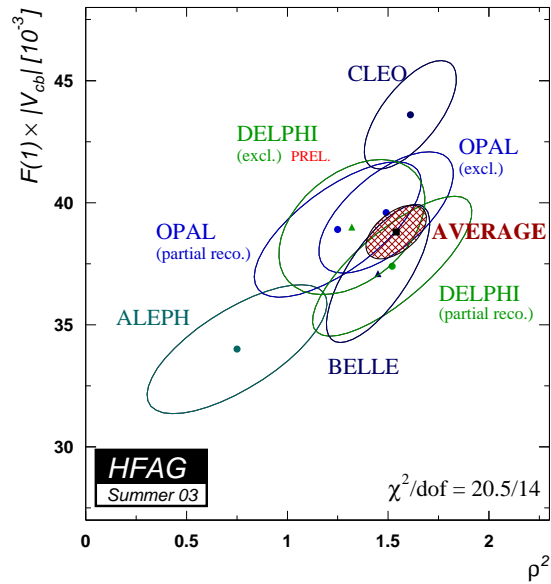}
\hss}
\caption{Correlation between ${\cal F}_{D^*}(1)|V_{cb}|$ and $\rho_{D^*}^2$ 
 measurements. The ellipses correspond to one sigma deviation.
 The world average has been performed by the Heavy Flavor Averaging Group~\cite{hfag},
 after rescaling the analyses to common input parameters. The label 'PREL.' refers to 
 the preliminary result of \cite{delphi_prel}.
 Only published results are taken into account in the average.}
\label{fig:elipses}
\end{figure}

Figure~\ref{fig:average} and~\ref{fig:elipses} shows the results and 
the world average of the different analyses, scaled to common inputs.   
Using the value of $F_{D^*}(1)$ given in Eq. (\ref{eq:f1}), it yields:  

$|V_{cb}|=(42.6\pm0.6(stat)\pm1.0(sys)\pm2.1(theo))\times10^{-3}$.
\vspace{-0.5cm}
\subsection{Uncertainties}
The dominant uncertainty on $|V_{cb}|$ is coming from theory. 
It is expected that lattice computations will improve the accuracy 
of the $F_{D^*}(1)$ value during the next years. 
The statistical error will also decrease below 1$\%$ as soon as
BABAR and BELLE analyze their total available statistics.
The other systematic uncertainties are coming from different sources. 
The most important contribution originates from errors 
correlated between the different experiments. For $|V_{cb}|$ these are the 
measurements of 
${\cal B}(b \to \overline {\rm B}^0_d)$ and 
${\cal B}(\Upsilon(4{\rm S}) \to \overline {\rm B}^0_d)$ rates, 
the ${\rm D}^{**}$ contribution and the branching fractions of the D
decay channels, whereas for $\rho^2_{D^*}$
the measurements of $R_1$ and $R_2$ form factor ratios are the dominant error
source. 

\section{$|V_{cb}|$ from $\overline{\rm B}~\to~{\rm D}\ell^-\overline\nu_\ell$}

$|V_{cb}|$ can also be extracted from the differential width:

\begin{equation}
\frac{d\Gamma(\overline {\rm B} \to {\rm D} \ell^- \overline \nu_\ell)}{dw}=
\frac{G_F^2|V_{cb}|^2}{48\pi^3}{{\cal K}_D(w){\cal G}^2_D(w)} 
\label{eq:widthd}
\end{equation} 

where, analogously to the
$\overline {\rm B} \to {\rm D}^* \ell^- \overline \nu_\ell$ case, 
${\cal K}_D(w)$ is a phase space function and ${\cal G}_D(w)$ is the form 
factor for the $\overline {\rm B} \to {\rm D}$ transition. 
The experimental difficulty in measuring $|V_{cb}|$ from this decay is coming 
from the large ${\rm D}$ contribution, from ${\rm D^*}$ decays, to the background,
especially at $w\simeq 1$, where the 
$\overline {\rm B} \to {\rm D} \ell^- \overline \nu_\ell$ decay rate is 
helicity suppressed. 
In addition, up to now, the theoretical control on ${\cal G}_D(1)$ is weaker than on 
${\cal F}_{D^*}(1)$ since the uncertainties coming from lattice computations 
have not been completely determined. Calculations using sum rules or the quark 
model find that ${\cal G}_D(1)$, unlike ${\cal F}_{D^*}(1)$ which benefits from 
the Luke's theorem~\cite{luke}, is affected by the first order 
$1/m_Q$ corrections. 
Averaging the ALEPH~\cite{aleph}, CLEO~\cite{cleod} and BELLE~\cite{belled}
measurements, gives~\cite{hfag}:
 
\begin{math}
\begin{array}{l}
{\cal G}_{D}(1)|V_{cb}|=(42.4\pm3.7)\times10^{-3};~\rho_{D}^2=1.14\pm0.16. 
\end{array}
\end{math}

\begin{figure}
\hbox to\hsize{\hss
\includegraphics[width=1.\hsize]{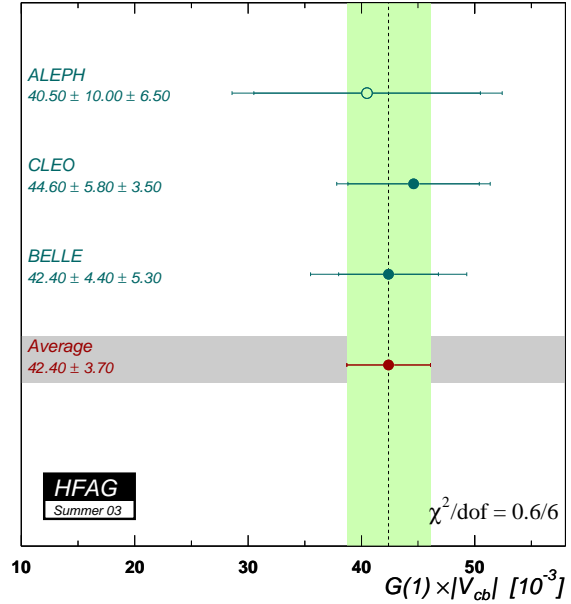}
\hss}
\caption{ The ${\cal G}_{D}(1)|V_{cb}|$ world average performed by the 
 Heavy Flavor Averaging Group \cite{hfag} after rescaling the analyses to 
 common input parameters.}
\label{fig:averaged}
\end{figure}

\begin{figure}
\hbox to\hsize{\hss
\includegraphics[width=1.\hsize]{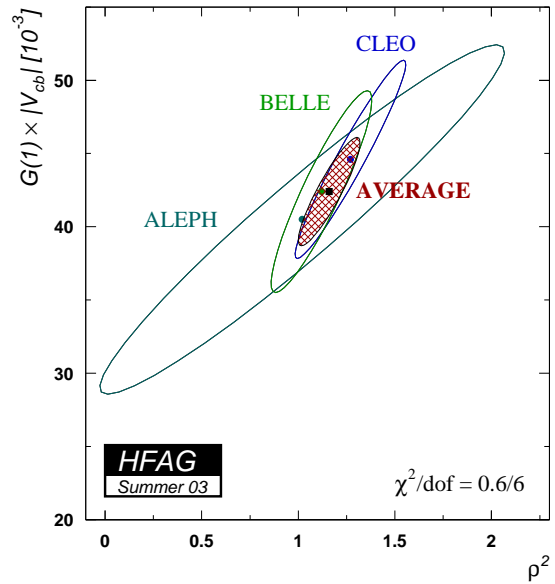}
\hss}
\caption{Correlation between ${\cal G}_{D}(1)|V_{cb}|$ and $\rho_{D}^2$ 
 measurements. The ellipses correspond to one sigma deviation.
 The world average has been performed by the Heavy Flavor Averaging Group~\cite{hfag},
 after rescaling the analyses to common input parameters.}
\label{fig:elipsesd}
\end{figure}

The contribution of each experiment, scaled to common inputs, can be seen 
in Figures \ref{fig:averaged} and \ref{fig:elipsesd}. 

Using ${\cal G}_{D}(1)=1.04\pm0.06$~\cite{ckm02}, it yields:

$|V_{cb}|=(40.8\pm3.6(exp)\pm2.3(theo))\times10^{-3}$.

which is compatible with the value obtained from  
$\overline {\rm B}~\to~{\rm D}^*\ell^-\overline\nu_\ell$ decays.
\section{Conclusions}

$|V_{cb}|$ has been exclusively measured by different experi\-ments using 
$\overline {\rm B}~\to~{\rm D}\ell^-\overline\nu_\ell$ and 
$\overline {\rm B}~\to~{\rm D}^*\ell^-\overline\nu_\ell$ decays. 

The averaged value obtained from $\overline {\rm B}~\to~{\rm D}^*\ell^-\overline\nu_\ell$
decays is:
\vspace{-0.3cm}
\begin{center}
$|V_{cb}|=(42.6\pm1.1(exp)\pm2.1(theo))\times10^{-3}$ 
\end{center}
\vspace{-0.3cm}
and from $\overline {\rm B}~\to~{\rm D}\ell^-\overline\nu_\ell$:
\vspace{-0.3cm}
\begin{center}
$|V_{cb}|=(40.8\pm3.6(exp)\pm2.3(theo))\times10^{-3}$.
\end{center}
\vspace{-0.3cm}
The dominant uncertainty is coming from the theoretical determination of the form 
factors at zero recoil, ${\cal F}_{D^*}(1)$ and ${\cal G}_D(1)$, which are expected 
to be improved 
by lattice calculations during the next few years. The experimental uncertainty is
limited by systematics due to input pa\-ra\-me\-ters such as 
${\cal B}(b \to \overline {\rm B}^0_d)$ and 
${\cal B}(\Upsilon(4{\rm S}) \to \overline {\rm B}^0_d)$ rates, 
the ${\rm D}^{**}$ contribution and the branching fractions of the D decay channels. 
These quantities have to be better determined to improve the $|V_{cb}|$ accuracy 
measured from exclusive decays.

\section*{Acknowledgements}
I would like to thank E. Barberio and U. Langenegger for computing the averages, 
and P. Roudeau for his invaluable co\-mments.

\end{document}